\documentstyle[aps,prl]{revtex}

\begin{document}

\title{Autonomous Energy Transducer: Proposition, Example, Basic Characteristics
}
\author{
        Naoko Nakagawa 
\\
        {\small \sl Department of Mathematical Sciences, Faculty of Science}\\
        {\small \sl Ibaraki University, Mito, Ibaraki 310-8512, JAPAN}\\
       and\\
        Kunihiko Kaneko\\
        {\small \sl Department of Pure and Applied Sciences}\\
        {\small \sl University of Tokyo, Komaba, Meguro-ku, Tokyo 153-8902, JAPAN}
\\}

\date{August 5, 2003}

\maketitle

\begin{abstract}
We propose a concept of autonomous energy transducer at a molecular scale,
where output is produced with small input energy, of 
the same order of the thermal energy,
without restriction of magnitude or timing of input, and without any
control after the input.
As an example that satisfies these requisites,
a dynamical systems model with several degrees of freedom is proposed,
which transduces input energy to output motion on the average.
It is shown that this transduction is robust and the
coupling between the input and output is generally loose.
How this transducer works is analyzed in terms of dynamical systems theory,
where chaotic dynamics of the internal degrees of freedom,
as well as duration of active state which is self-organized with the energy
flow, is essential.
We also discuss possible relationships to enzyme dynamics or protein motors.
\end{abstract}

\section{Introduction}

How energy is converted from one form to another form has been
extensively studied in these days.  In biochemical reaction, some molecules are known
to work for ``energy transducer".  Such energy transduction is
important in enzyme function, molecular motor, and so forth,
while recent progress in
nanotechnology makes us possible even to design a microscopic machine 
that works as such energy transducer.

The question whether such molecular machine functions in the same way as
a macroscopic machine was first addressed up by Oosawa \cite{Oosawa_Hayashi,Oosawa} as a dichotomy
between tight and loose coupling  
between input and output.  In the tight coupling, 
chemical energy is converted into mechanical work 
via definitely scheduled successive conformational changes of molecules, 
and at each stage the conformational change has one-to-one correspondence
to a mechanical work.
This mechanism is similar to most of our macroscopic machine.  Consider
a typical machine with gears.  It converts 
some form of energy to work via several rigid gears, 
following external operations.  Each gear transforms one kind of
motion to another directly and instantaneously, where
fluctuations in the transformation process are negligible.

In contrast, in the loose coupling mechanism, the manner or the route
to establish energy conversion 
does fluctuate by each event, and the energy is not necessarily 
converted all in once at each stage.
Here, the amount of output work is not precisely specified, 
but varies by each event. By repeating trials, the output work
extracted from the same amount of input energy (e.g., by a single reaction 
from ATP to ADP) is distributed with some fluctuations.
Oosawa proposed that this
loose coupling is a general property of a system
that {\it robustly works} under large thermal fluctuations of 
the configuration of molecules even though the input energy 
is of the same order with the energy of thermal fluctuations.

This problem is first posed in the study of molecular motors, where conversion 
from chemical energy to directional mechanical work within a fluctuating
environment is studied.  Indeed, there are some experimental results
suggesting loose coupling mechanisms.
For instance, the energy is stored in a molecule over an anomalously long time 
between the chemical reaction event and emergence of 
directional motion \cite{Ishijima_Yanagida},
and multiple steps of conversion to work are repeated 
during molecular processes within
one cycle of the chemical reaction \cite{Kitamura_Tokunaga}.
There the resulting output steps is widely distributed
as is expected in loose coupling mechanism.

Whether the energy conversion of a biomotor is loose or tight
may depend on detailed molecular structure.  Still,
the question how the energy conversion in loose coupling mechanism 
works is quite general,
and can be discussed in any interacting molecules.
One possible mechanism for the
loose coupling was proposed as thermal ratchet at a {\em statistical} level 
\cite{Feynmann,Vale_Oosawa,Magnasco,Prost}, although {\em dynamical} process for the
loose energy conversion is not yet clarified.
Little is known what types of loose conversion are possible, or
in what condition and by what mechanism the loose coupling functions.   

In the present paper, 
we propose a scheme of {\em dynamical energy transduction
of autonomous regulation} in highly fluctuating configurations, 
by introducing a system consisting of interacting molecules 
with internal dynamics.
The specific choice of a model we adopt for this energy conversion
is inspired  by 
some experiments suggesting  oscillatory molecular dynamics 
at a time scale of molecular work \cite{Hatori_Honda2,Shingyoji},
whereas we do not pursue to make a direct correspondence with the experiment.  Rather,
we adopt the model, expecting that the proposed mechanism itself is rather general.
By setting up an example of a dynamical systems model that works as a loose coupling,
we show that energy transduction is possible within fluctuating environment.
By analyzing the dynamical process of this energy transduction, 
we find some characteristic features of the dynamics
that makes the coupling in this scheme generally loose.  
Robust transduction of energy within fluctuating configuration is discussed
from the viewpoint of dynamical systems.  
The proposed mechanism may also shed new light on the function of enzyme 
in chemical reaction, design of a nano-machine that
works at a molecular scale, and biomotors.

\section{Definition of Autonomous Energy Transducer at Molecular Scale}

To study this problem, we start by defining a machine that works 
at a molecular scale, with which we are concerned here. 
The requisite here is 
essential to a system that works
with large fluctuations of configurations in molecules.

Consider a system composed of several degrees of freedom.  `Input' is
injected into the system through some restricted degrees of freedom of it, that are
pre-defined.
Then, as a reaction to the input, there arises dynamical change in the system,
which leads to `output'
as a typical motion of another set of restricted degrees of freedom.
Here, the degrees of freedom for the input and for the output  are distinct.
Statistically, the output brings directional flow or mechanical work,
although individual events of the output would be rather probabilistic.
The dynamics of the system is given by the interaction among the
degrees of the freedom, as well as  noise from the heat bath.
Through the dynamics, input energy is transformed to (stochastic) output motion,
that gives an output work statistically.
This energy transducer
works with large fluctuations of configurations of the system,
without any external control,
once the energy is injected as input.
Accordingly we postulate the following
properties:

\begin{itemize}

\item[(i)] Details in input event are not controlled.  
Neither the timing of the injection
with regards to the molecule fluctuations, nor the direction of the input 
motion is specified precisely.  Even the magnitude of the input is
not restricted. 

\item[(ii)] Once the input is injected, no external control is allowed, 
and the process
is governed by a given rule for temporal evolution, while 
fluctuations at a molecular scale are inevitable.

\item[(iii)] In spite of this uncontrolled condition, 
mechanical work or directional flow can be extracted from the output motion,
on the average.  The system adapts itself to produce output,
against the change of the input timing, direction, or magnitude.

\end{itemize}

Then, we address the question how such system is possible.
In the present paper, we provide a
specific example of a system with such loose-coupling,
and discuss its characteristic features which
realize the above three conditions.
The characteristic features of dynamical process are summarized
as follows:

After the energy is given as the input, 
the input energy is sustained with very weak dissipation for some time,
and is not directly transduced to the output.
This suppression of energy flow into other modes comes from some restriction
to the interaction among the system. 
On the one hand, this restriction 
makes a change of 
molecular dynamics from the fluctuating motion to the
directional motion later.

During the time while the energy is stored, detailed information of the
input (such as the direction or the timing),
except the energy value, is lost due to noisy chaotic dynamics.
This makes the output distributed.
In other words, the input-output relationship is loose,
in the sense that the output (i.e., displacement or extracted work) is
distributed around its mean.  
There is a rather general relationship
between the variance and the mean of the outputs, as will be shown.

By studying a dynamical  model with several degrees of freedom,
we have confirmed the above features,
when the model satisfies the requisite (i)-(iii),
to extract mechanical work 
from the energy flow within highly fluctuating configurations.
On the other hand, no substantial work is extracted from the output, when
these features are not observed in the models.
These results imply that the
autonomous energy transduction works as a loose coupling mechanism.

\section{Model}

As a specific system to satisfy the condition (i)-(iii), we consider the
following model
\footnote{The present model is originally inspired
by the study of motor protein (e.g., myosin) that moves along the chain of 
the rail protein, actin.} \cite{Nakagawa_Kaneko4}.
The motor part consists of a `head', of position $x_h$, and
an internal `pendulum' represented by $\theta$.
The motor interacts with rail, represented by a
one-dimensional chain of lattice sites positioned at $x_i$ with index $i$.
The internal pendulum is excited when energy is applied to the system.
Every degree of freedom, except for $\theta$, is in contact with a heat bath,
generating random fluctuations described by a Langevin equation
with damping.
The interaction potential $V(x_h-x_i,\theta)$
between the chain and the head is spatially asymmetric and its form
depends on the angle of the pendulum (see Fig.\ref{fig:model}).
The periodic lattice is adopted as the chain,
to study directional motion in an asymmetric periodic potential,
as is often studied in the study of thermal ratchets.
Every degree of freedom, except for the internal pendulum, 
is in contact with the  heat bath,
generating random fluctuations described by a Langevin equation.

The equations of motion for this system are chosen as
\begin{eqnarray}
m_c \ddot x_i &=& -\gamma \dot x_i +\sqrt{2\gamma T}\xi_i(t)\nonumber\\
&-&K_c\left\{ (x_i-iL)+(x_i-\frac{x_{i-1}+x_{i+1}}{2})\right\}
-\frac{\partial V(x_h-x_i,\theta)}{\partial x_i}, \nonumber\\
m_h \ddot x_h &=& -\gamma \dot x_h +\sqrt{2\gamma T}\xi_h(t)
-\sum_i\frac{\partial V(x_h-x_i,\theta)}{\partial x_h}, 
\label{eqn:model}\\
m_{\theta} \ddot \theta &=& -\sum_i\frac{\partial V(x_h-x_i,\theta)}{\partial \theta},\nonumber
\end{eqnarray}
where $T$ is the temperature, $\gamma$ is a friction coefficient 
and $\xi_{\alpha}(t)$ represents Gaussian white noise.
Here, we use the units Boltzmann constant $k_B=1$.
$K_c$ and $L$ are the spring constant and the natural interval 
between two neighboring lattice sites in the chain.
To observe directional motion of the head, the chain is also connected to 
a fixed ground via a spring with a constant $K_c$. Here,
$m_c$, $m_h$ and $m_\theta$ are mass of the respective degrees of freedom,
$m_c=m_h=1$ and $m_{\theta}=O(0.01)$.
The inertia is not ignored, as small friction coefficient $\gamma$  is adopted.
Although this could be reasonable for the application to enzymatic reaction or
to a design of nano-machine, one might think that
biomotors should be treated as overdamped systems, considering
that for small object that moves with a slow time scale, water is 
highly viscous fluid.
Here, we introduce inertia term, because several experimental results
by Yanagida's group may not necessarily match with this standard view
\cite{Ishijima_Yanagida,Kitamura_Tokunaga,Shingyoji,Yanagida_Oosawa,Ishijima_Yanagida0,Nishiyama_Higuchi,Wazawa_Yanagida,Yokota_Yanagida}.
Also,  each variable in our model
does not necessarily correspond to the atomic configuration  of a protein.
Rather it is a mesoscopic variable, for which effective inertia may be  relevant to long-term
storage of energy in dynamics of proteins,
and allow for oscillatory dynamics
among interacting degrees of freedom.
We will come back to this 'inertia problem' in \S 8 again.

The potential form is asymmetric in space as shown in Fig.1,
where the characteristic decay length of the interaction is set 
at a smaller value than $L$, so that the interaction is
confined mostly to the nearest lattice sites. 
Here we adopt the following potential form,
$
V(\Delta x,\theta)= V_{x}+V_{\theta}
$
with
$
V_x= K_h{\tanh(p(\Delta x -r))}/{ \cosh(d\Delta x)},
$
and 
$
V_{\theta}=K_h(1-\cos\theta)/2{ \cosh(d\Delta x)}.
$
Here the parameters $p$ and $r$ determine the degree of asymmetry, while
$K_h$ and $d^{-1}$ give the coupling  strength and decay length of the interaction, 
respectively.
Specific choice of this form is not important.
We have simulated our model choosing several other potential forms with asymmetry, 
for instance
$
V(\Delta x,\theta)= K_h(1-\cos\theta)/(\exp(p\Delta x)+\exp(-d\Delta x)),
$
and obtained qualitatively the same results with regards to the directional motion.
In this paper, the parameters for the potential are fixed at
$p=10$, $r=0.3$, $K_h=0.2$ and $d^{-1}=0.25$. 

The pendulum mode without direct coupling to the
heat bath is adopted here just as one representation of
long-term storage of energy \cite{Nakagawa_Kaneko} experimentally suggested
in molecular motors \cite{Ishijima_Yanagida}. As long as there exists 
such slow relaxation mode, our results follow, and this specific choice 
is {\em just one example} for it. 
Any mode realizing slow relaxation 
can be adopted instead.

\section{Output directional motion }

Once a sufficient energy is injected to the pendulum to drive 
the system out of equilibrium,
we have observed that the head moves a few steps in one direction,
on its way in the  relaxation  to thermal equilibrium \cite{Nakagawa_Kaneko4}.
An example of the time course of the head motion is given 
in Fig.\ref{fig:jikeiretsu}.
How many steps the head moves after the energy injection depends on
each run with different seed for the random force, even though the values of
parameters including $E_0$ are kept constant.

In Fig. \ref{fig:temperature}(a), 
the average step size is plotted as a function of temperature ($1/T$).
The mean step size only weakly depends on the temperature for small $T$.
It shows only slight decrease by decreasing the temperature.
This is in strong contrast with the thermally activated process,
in which the rate of crossing the potential barrier (for the head to shift)
is given by $\exp(-U_0/T)$ with an activation energy $U_0$,
according to Arrhenius law.
At thermal equilibrium, indeed, the head shows such thermally activated motion.
The resulting diffusion rate of the head Brownian motion obeys
this exponential form as shown in Fig.\ref{fig:temperature}(b), whereas
the values of $U_0$ agree with the activation energy 
derived analytically \cite{Nakagawa_Kaneko6}.

The dependence in Fig.\ref{fig:temperature}(a) is distinct from
such dependence, and
the motion of the head
is not thermally activated.  This is natural, since
the energy for the directional motion is extracted from the
injected energy $E_0$.
As a result, the head motion possesses different properties from the thermally 
activated motion as follows.
For a fixed temperature, the average step of the head increases monotonically 
(almost linearly) with $E_0$, when the injected energy $E_0$ is 
larger than a threshold (see Fig.\ref{fig:injected_energy}(a)).  
This threshold is not equal to $U_0$ for the Brownian motion,
but shows the  following dependence upon temperature:
For higher temperature, this threshold energy goes to zero, and
any small injected energy can lead to the directional motion on the average,
whereas, with decreasing temperature, the threshold energy tends to converge 
to a certain value around $0.3$. 
This result suggests that there is a certain minimum energy required 
for the head to shift to the next lattice site.  

The temperature dependence suggests that the head motion is not thermally 
activated.  Rather, it seems that
the motion takes place when a suitable temporal correlation exists
between the head motion and thermal fluctuations of the chain.
It should also be noted that the directional motion is clearly
observed even by the injection of the energy of $O(10)$ times of
$k_BT$.  For example, the average motion is about 2 steps, for the
injection of the energy of $20k_BT$, for $T=0.02$.

The linear increase of the step size with $E_0$ implies that
the injected energy is not wasted even if it is larger than the required 
threshold value for the directional motion of the head.  
This implies that the injected energy is converted to the 
directional motion not at once but step by step with the successive step motion of the head.
This feature is revisited in Section 6.

To make this successive energy conversion possible, it is important
that the chain be sufficiently flexible.  
Indeed, as is shown in Fig.\ref{fig:stiffness}, if the
spring constant $K_c$ is too large, directional motion
of the head is not observed.
The directional motion is not possible, if the chain is too tight, 
or too flexible.  
The average step size has a peak around $K_c \sim 0.5$.  
This suggests that a flexible motion that appears under a proper stiffness,
plays an important role to produce the directional motion 
as an output of the energy transduction.
Indeed, around this parameter value, there is a
transition in the dynamics of the head and neighboring few lattice points,
to show correlated motion, and accordingly
a rather long-range correlation appears there\cite{Nakagawa_Kaneko6}.

\section{Characteristics of Dynamic Energy Transduction with Loose Coupling}

Now we discuss how the above directional motion is achieved in the term of dynamical systems.
We note that the present mechanism works only as a loose coupling mechanism
in this section. 

First, let us discuss how the system `forgets' initial conditions 
and how broad distribution of output step motion appears.
Before the directional motion of the head, the motion of 
the triplet \{head, pendulum, rail \} shows a weakly chaotic motion,
with which the input information (e.g., the configuration of the system
when the input is applied) is lost.  
The chaotic motion here only works as weak perturbation to the two-body
problems consisting of the interactions between the head and the pendulum and
that  between the head and the adjacent lattice site.
On the other hand, when the head exhibits a step motion,
there is strong interaction between the head and the lattice sites,
that lead to a stronger chaotic instability  
arising from three-body or
five-body (including the neighboring lattice points) motions \cite{Nakagawa_Kaneko6}.

The change of the degree of instability is demonstrated in Fig.\ref{fig:difference}, 
where the evolution of difference between two close orbits is plotted.  
As shown, there are two rates of exponential divergence, one 
that continues during the energy storage, and
the other with the flow of energy from the pendulum part.
The former is given by perturbations to the two-body motions, and the latter
is induced by strong interaction with the three- or five-body motion.  
The orbital instability here is examined by
taking two identical initial configurations, and applying identical input energy and
identical random noise (i.e., sequence of random number). 
A slight disturbance is added to the reference system 
at the timing of energy injection.
The evolution of these two systems with tiny difference in configurations
results in quite different number of output steps 
(see Fig.\ref{fig:difference}(a)).  
Then, it is impossible to predict or control the number and the direction 
of steps by the configuration of the system at the excitation.
On the other hand, the 
directional motion appears rather independently of the configuration
and, therefore the energy conversion of the system is  robust against
thermal noise, and disturbances in the input event.

One consequence of these chaotic dynamics together with
the influence of the heat bath is a {\it loose} relationship between
the input and the output.  In fact, the output number of steps
shows a rather  broad distribution, as  shown in Fig.\ref{fig:distribution},
where one step means the displacement of the head for one lattice site 
along the chain.  
The distribution was obtained by taking $1000$ samples
with arbitrarily chosen configurations  
at the moment of energy injection,
while satisfying thermal equilibrium,
and also by taking different
random sequences $\xi_{\alpha}(t)$. 
The resulting distribution is  rather broad, 
although the input energy  $E_0$ is identical for
all the samples.
The difference in the output step comes mainly from
intrinsic chaotic dynamics of the system mentioned above.

In the distribution of the number of steps, we note that
at low temperature the mean displacement is almost equal to
its variance (see Fig.\ref{fig:distribution}(b)).  
If the step motion is a result of 
rare stochastic events, one could expect Poisson distribution.
Indeed, if one disregards the small tail 
of the distribution at negative steps, the obtained distribution is
not far from Poissonian, for low temperature.
Even for high temperature, the linear relationship between the mean and
variance is still valid with an offset depending on the temperature.
This Poisson-type distribution suggests that random process
underlies the output head motion, as expected from the above chaotic dynamics.

The relationship in Fig.\ref{fig:distribution}(b) indicates that to obtain
sufficient output amount
in directional motion, large fluctuation is inevitable.
Indeed, it is not possible, at least in the present model,
to make the step distribution sharp while keeping sufficient directional 
motion of the head,  even if the parameter values are changed.
For example, one could
make our model rigid, 
by increasing  stiffness $K_c$ in the rail and suppressing the fluctuations of the configuration.  
Then, the directional motion is highly suppressed, 
as already shown in Fig.\ref{fig:stiffness}.
The average displacement as a function of the stiffness has a maximum
around $K_c \sim 0.5$, where the fluctuation of the displacement of rail elements is rather large.
On the other hand, for $K_c=1.3$, i.e.,
for a rigid spring of the rail, the average
displacement is only 2\% of the maximum. 
There, most of the input energy then is wasted.
Accordingly, to achieve larger directional steps without wasting 
input energy, the step distribution has to be broad.
This demonstrates the relevance of 
loose coupling mechanism to energy transduction.

This relevance of loose coupling is 
expected to be general under our postulates for the molecular transducer.
Note that rigid process in tight coupling mechanism can transfer energy 
very well when configurations of every two parts
between which the energy is transferred is precisely determined,
but the efficiency will go down  drastically if this precise condition for the configurations 
is not satisfied.  In the molecular energy transducer we are concerned,
such precise control of configurations of molecules is impossible.
Furthermore, as the detailed information on the input is lost, control of the motion 
gets more difficult.
Hence, the tight coupling mechanism is difficult to keep high efficiency
for a molecular transducer.

On the other hand, in the present 
loose coupling mechanism for the energy transduction,
there exists chaotic motion.  
Then, the efficiency could be much lower than
the maximum efficiency
achieved by tuning initial condition in the tight coupling.
However, the probability to achieve 
`good' configurations starting from any
initial conditions or from a variety of boundary conditions remains to be large.  
Hence energy transduction in present loose coupling mechanism 
works over a wide range of initial conditions, and
therefore is robust.

\section{Characteristics of Energy Flow leading to Directional Motion}

In spite of chaotic motion mentioned above, the output should be
directional, in order to extract some work from the system.  Here, recall that
there is some time lag between the injection of
energy and the emergence of the step motion, during which chaotic motion is maintained.  
Even after the stronger chaos appears with the flow of energy 
from the pendulum, there is a certain time lag 
before the emergence of the step motion (see Fig.\ref{fig:difference}).
During such time lag, the energy may be sustained 
or may be used degree by degree to bring about the step motion.
In this section, we examine the energy flow in detail during
this `active state' preceding to the directional motion. 
The duration of active state is suggested to be important 
to realize the directional motion 
under large fluctuations of the configurations of molecules.

Note that the motion of the head to the neighboring lattice site is
possible only if some condition for the configuration among the pendulum,
head and the adjacent lattice sites is
satisfied.  In Fig.\ref{fig:jikeiretsu}, we have plotted the time
course of energy flow to the chain from the head-pendulum part.  Here, the 
time series for the kinetic energy
of each lattice site is plotted together with that of the head.
As shown, the energy flow from the head
to farther lattice site is suppressed,
before the step motion occurs.  This restriction
in the energy flow is generally observed whenever the directional motion 
is observed.
On the other hand, there is a continuous gradual flow 
when the step motion is not possible.  

To study the  relevance of this restriction in the energy dissipation,
we have computed energy storage by measuring the following two quantities;
\begin{eqnarray}
W_{in}(t) &\equiv& -\sum_{j=i_0-1}^{i_0+1}\int_{0}^{t} 
{\dot x_j}\frac{\partial V_{\theta}}{\partial x_j}dt,\\
W_{out}(t) &\equiv& -\int_{0}^{t} 
{\dot x_{{i_0}-2}}\left\{
\frac{\partial V}{\partial x_{{i_0}-2}}-\frac{K_c}{2}(x_{i_0-1}-x_{i_0-2}-L) \right\}dt \nonumber\\
 && -\int_{0}^{t} 
{\dot x_{{i_0}+2}}\left\{
\frac{\partial V}{\partial x_{{i_0}+2}}-\frac{K_c}{2}(x_{i_0+1}-x_{i_0+2}-L) \right\}dt.
\end{eqnarray}
$W_{in}(t)$ is the work from the pendulum to the head and the 
lattice, up to time $t$ since the energy injection,
and $W_{out}(t)$ is the work from the three lattice sites adjacent to
the head, to farther lattice sites, again up to the time $t$.  
Here we call the head, pendulum,
and the adjacent three lattice sites as "active part", and other
lattice sites of the chain  as "residual chain part".
By  adopting these terms,
$W_{out}(t)$ is the
energy dissipated to the residual chain part from the active part.
In the present set of simulations for the measurement of $W_{in}$ and $W_{out}$
( and only in this set of simulations),
the contact with the heat bath to the active part 
is eliminated, to distinguish clearly
the stored energy from the wasted work to other lattice sites.

A time course of $W_{in}$ and $W_{out}$ is shown in Figs.\ref{fig:flow}.
In a certain range of parameters where the step motion frequently occurs,
dissipation from the active part is delayed for some time after
the energy flow from the pendulum into the system (Fig.\ref{fig:flow}(a)).
During the interval for this delay, the energy is sustained in the active part
(i.e., at the head and the adjacent lattice sites) for a while.
As is shown in the left figure of Fig.\ref{fig:flow}(a), the sustained energy is dissipated 
to the residual chain, together with the event of the step motion.
(Note that steep changes in $W_{in}$ and $W_{out}$
after the step motion is an  artifact since
the computation here is based on the active part consisting of
the head that is adjacent with the site $i_0$,
and loses its meaning after the head steps from the $i_0$th site.)
For the parameters with much smaller directional motion, 
the increase of $W_{in}$ and $W_{out}$ begins simultaneously with
the energy flow from the pendulum to the system
(Fig.\ref{fig:flow}(b)).

The mean profile of the energy flow from the active part 
is given by the time course of $d\langle W_{out}\rangle /dt$, where
the temporal ensemble average $\langle \cdot\rangle$ is computed by setting
the origin of $time$ as the time of the energy injection,
since, for the parameters adopted here,  
the energy flows rapidly (in less than $10$ time units) from the pendulum 
to the other degrees of freedom.
The rate $d \langle W_{in} \rangle /dt$ takes a large value 
only in the initial stage ($time<10$) of the relaxation process.
The ensemble average at time $t$
is computed only over the samples in which the head stays at the
site $i_0$ up to time $t$.  Accordingly,
the number of samples decreases with the increase of the time.

As is seen in Fig.\ref{fig:mean_flow}(a), the out-flow rate 
$d\langle W_{out}\rangle /dt$ of the energy is kept small at the early stage 
(up to $time\simeq 100$) of the relaxation, for $K_c=0.5$.
This is in strong contrast with the temporal profile
for larger values of $K_c$ (e.g. $=1.3$).
There, the in-flow rate $d\langle W_{in}\rangle /dt$ to the active part
is approximately $0.01$ in maximum for both the two cases.
Then, the active part in the flexible system does not respond instantaneously 
to the sudden flow of energy, and therefore
the energy is sustained in the active part without dissipation into
the residual chain for a while.
This sustained energy up to $t$ is estimated by $W_{in}(t)-W_{out}(t)$.
The ensemble average of the time course of this energy 
is given in Fig.\ref{fig:mean_flow}(b).
The energy is sustained up to the time $t \approx 100$ when $K_c=0.5$
and dissipated more quickly when $K_c=1.3$.

If the injected energy were dissipated completely into a large number 
of degrees of freedom,
the conversion of energy into mechanical work
would suffer a rather large loss.
Contrastingly, the motion generated by our mechanism remains
confined within just a few degrees of freedom where
some correlation among degrees of freedom is maintained.
Supplied energy to the pendulum is not diffused
as heat.  As a result, the conversion is efficient.
The restriction of energy transfer to the farther lattice sites 
comes from the correlated motion among the pendulum, head, and 
the lattice sites \cite{Nakagawa_Kaneko4,Nakagawa_Kaneko6}.

The effective rate for the sustained energy in the active part
is estimated by 
$\langle W_{in}(t_w)-W_{out}(t_w)\rangle/\langle W_{in}(t_w)\rangle$ 
by suitably choosing $t_w$.
(As $t_w$ is increased to infinity, these quantities go to zero, of course).
By choosing the time $t_w$ as an average waiting time
for the step motion, one can estimate the sustained energy
within the time scale for the step motion.
As shown in Fig.\ref{fig:maintained_energy} approximately $80 \%$ 
of the energy flow is sustained in the active part for
$K_c \approx 0.5$, where the average steps in directional motion
becomes maximum.
The mean waiting time $t_w$ is almost $100$ for parameter values 
adapted in Fig.\ref{fig:maintained_energy}, which is much shorter
than the time for the complete dissipation of the energy. 

Now it is shown that
the combination of chaotic motion (for the loss of details of input) 
and the restriction of dissipation (that leads to directional motion) 
gives a basis of our 
mechanism how the energy is transduced in a microscopic
system under large fluctuations of their configurations.
Due to the chaotic motion, if
the head shifts to the next site by crossing the energy barrier or not
is probabilistic.
Here, the active part 
can sustain the excess energy effectively. Therefore, the head comes close to
the barrier several times even if it fails in the first trial.
Hence the motion to cross the barrier can
be robust and adaptive.

Revisiting Fig.\ref{fig:jikeiretsu}, it is noted that the amount of 
the stored energy $E_{\theta}$ in the pendulum decreases step by step,
corresponding to the step motion of the head.
The energy once extracted from the pendulum is partly restored 
after the completion of the step motion.
The fact that the energy is restored to the pendulum 
allows the delayed use for the multiple step motion.
Thus, in our mechanism the energy is used step by step.

To confirm the delayed use in multiple steps, 
we have measured $W_{in}$, by restricting only to such case that 
multiple steps are observed (Fig.\ref{fig:degree_by_degree}).
Here we have computed $W_{in}$ up to the time of the first,
second, and third steps respectively.  These measurements
show that the extracted energy $W_{in}$ from the pendulum 
up to the $n$-step motion increases with $n$.
These results imply that the multiple step motion is supported 
by the delayed step-by-step use of the stored energy.  This is why
output motion with more and more steps is possible
as the injected energy is increased.

In Fig.\ref{fig:degree_by_degree}, one may note that the amount of 
the available energy in the pendulum for the next steps
is smaller than that for the previous step.
Indeed, the extracted energy in Fig.\ref{fig:degree_by_degree} is estimated
to take a rather large value due to the timing to measure it, i.e.,
$W_{in}$ is measured at the timing 
when the head crosses the energy barrier to move to the next site.
Then, the  energy stored in the pendulum is expected to be larger than 
expected.  This is because some part of the energy is 
restored to the pendulum when the head approaches the 
bottom of the potential for the next site, 
as is seen in several rises of $E_{\theta}$ after the first step
motion shown in Fig.\ref{fig:jikeiretsu}.

\section{Discussion}

In the paper, we have proposed a concept of autonomous energy transducer 
at a molecular scale.
In this transducer, we have set the constraints (i)-(iii) in Section 2:
Output is produced, even though
input energy is of the same order of the
thermal energy,
and even though the magnitude or the timing of input is not specified, nor any
control after the input is introduced.  
This requisite is postulated so that
the function of the autonomous energy transducers is robust
at a molecular scale.

We have shown a possible scheme that satisfies this requisite,
by providing a simple example, with a dynamical system of several degrees of freedom.
In the model, the energy transducer is robust by the
following two properties:
The first one is chaotic dynamics.
The chaotic dynamics amplifies difference of initial conditions by each  input
and thus detailed condition on the input event is lost.
Chaotic dynamics  also brings mixing of orbits,
and each orbit from different initial condition has a chance to visit a certain part of phase space 
that is necessary to generate directional motion.
Hence, an output motion can be obtained on the average,
independently of details of the input.

The second property is energy ``storage'' at ``active part'' in the system.
The active part consists of several degrees of freedom, 
and is self-organized by absorbing energy 
from the input degrees of freedom.  This active part is 
sustained over rather long time interval,
with very weakly dissipating the energy.
During this long time span,  each orbit has more chances  to visit
the part of phase space inducing directional motion. Then, the ratio
of the events with the directional output motion is increased.

In addition, a ``storage unit", like a pendulum, 
enhances the potentiality of autonomous regulation of the system.
The dynamic linking between the storage and the active parts 
makes restorage of the extracted energy possible,
which makes the delayed use for multiple steps easier.
Such regulation would not be possible if the energy were simply 
provided to the active part, as assumed for a single conformational change 
per one chemical process.

As a consequence of our scheme of energy transduction,
the output has large fluctuations by each event.  
There exists variance of the same order with the mean of the output.  
Hence input-output relationship is not one-to-one, but the output
from the same input is distributed.  Now, a mechanism is provided for
a molecular machine with loose coupling between the input and the output, 
as was originally proposed by Oosawa.

For the autonomous energy transducer,
the output  has to be  extracted without specifying input conditions, even
under the fluctuations of the same order of the input.  
Hence, in order for the transducer to work at a molecular scale,
several degrees of freedom are necessary to exhibit flexible change.
The input energy is first
stored into some of these degrees of freedom, and then is used step by step.
Accordingly, the output is not directly coupled to the input.
In contrast with a tight-coupling machine consisting of "rigid gears",
the machine with this  loose coupling is flexible.  

After the output is extracted from the transducer,  
a stationary final state is restored, that is identical with
the initial state before the input, except the change for the output
(i.e., translational motion).
In the model we presented, 
such stationary state is represented by a combination of stable two-body motions,
that are almost decoupled each other.
During the transduction process, however, several degrees of freedom in the
internal states are coupled.
This mode coupling leads to flexible dynamics on the one hand, 
but also brings about the directional motion, on the other hand.
Relevance of the change of effective degrees of freedom to energy transduction
has also been discussed in relationship with chaotic itinerancy \cite{Nakagawa_Kaneko6}.

Although we presented our scheme by a specific model with a head and a rail,
the mechanism proposed here is expected to be rather general.  
What we have required is just a dynamical system with several 
degrees of freedom, including those for input and output.
There, the system changes effective degrees of freedom 
through the internal dynamics in the system.
We note here that each degree of freedom we used in the model
does not necessarily have to correspond to an atomic motion.  Rather, it may
represent a collective variable consisting of a large number of atoms.
It should be also noted that the present scheme for the energy transduction
is not necessarily restricted to a Langevin system with weak damping, 
but it is hoped that a model with overdamped dynamics will be constructed
that realizes energy transduction with loose coupling 
by the present scheme.
In this sense, the concept of autonomous transducer at a molecular scale, which
we have proposed here, would be general and applied widely.

The importance of loose coupling mechanism lies in  its adaptability to different
conditions.  Although the motion is not precisely optimized with regards to the
efficiency of the transduction, instead, the present transducer works
even if external conditions are changed.  

As an extreme case, consider such external condition that makes
directional motion much harder.
For example, by adding a load to the transducer, it is expected that
the output motion is suppressed.  Then the question is raised
if the fluctuation in the output may also be decreased.
If the answer is in the affirmative, the coupling between input and output will be more tight,
as added load is increased.
In other words, a transducer with loose coupling can show tight coupling
depending on external condition.  On the other hand, a system designed to have a tight-coupling 
cannot work as a machine with a loose-coupling, under the change of external condition.
It just does not work when the condition is changed.

To check this plasticity of energy transducer with loose-coupling,
we have carried out a numerical experiment of our model
by adding a load.
In order to study the effect of a load, we make a slight modification of the present model;
We reverse a condition for fixing the total system, i.e.,
the head is fixed (e.g., by attached to a glass plate),
and the rail (chain) is set free, and its center of mass can move.
Due to the friction from the heat bath, the rail itself imposes 
a certain amount of load to the head. The amount of the load 
increases with the length of the rail given by $N_c$.

In Fig.\ref{fig:load-distribution}, the step distributions are displayed
for two different lengths of chain, ($N_c=5$ and $200$).
When the length of the chain is short, the mean step size is multiple 
that is distributed rather broadly by samples, 
similarly to the original system (refer Fig.\ref{fig:distribution}).
On the other hand, for a sufficiently long chain, the mean is still kept 
positive but becomes smaller in amount. Moreover, the distribution gets
much sharper. The fluctuation of the output is very small, i,e, the output is
almost constant, as is expected in the tight coupling mechanism.
Such trend is also observed in the experiments of molecular motors \cite{Ishijima_Yanagida0}.

\section{Possible Application of  Autonomous Energy Transducer}

A good candidate of the autonomous energy transducer is protein.
In applying the present idea to proteins, it is interesting to recall that several
reports show that protein is not rigid, but flexible.
Actin, a rail protein, is known to be not so stiff, as assumed in the chain
of our model.  Some proteins are known to take several forms and spontaneously switch
over these forms under thermal energy.

Recently, there are several reports on single-molecule protein dynamics,
by adopting fluorescent techniques.  Some enzymes show long-term relaxation of a time scale of
m-sec to seconds\cite{Lu,Edman}.  In particular, oscillatory dynamics  with order of seconds
are observed in the conformational change  of
single enzyme molecules, which suggest cyclic dynamics over transient
states\cite{Edman}.  
Anomalously long-term energy storage is also observed
by Ishijima et al \cite{Ishijima_Yanagida}.
There the chemical energy from ATP hydrolysis is stored in the motor
for 0.1-1sec and then the energy is used for the directional motion gradually.
The energy is used step-by-step for multiple steps \cite{Kitamura_Tokunaga}.
Oscillatory dynamics are also suggested in Dynein motor \cite{Shingyoji}. 
Summing up, several
experiments support long-term dynamics in the relaxation of protein
after excitations, as assumed in the dynamics of our model.

If the output is simply provided by chemical reaction, for instance by
hydrolysis of ATP, 
such dynamical behavior would not be possible.
For the dynamics, the proteins in concern should possess a ``storage unit" 
between the domain for chemical reaction and the active part.
In the present model, 
the storage unit is a simple pendulum and is not directly contacted with the heat bath. 
This is just an idealization,  and the storage unit need not  be so simple.
Also, the obtained results would be valid qualitatively even though a weak 
contact with the heat bath is added.

From our result, it is postulated that 
the storage unit should interact with the active part bi-directionally, 
and the dissipation of energy from the unit is rather weak.
Then, multiple step-by-step motion with restorage of the energy
is possible.
We hope that the storage unit that satisfies the above requisite 
is discovered in motor proteins.

Although the purpose of our paper is to propose a new general mechanism for
loose coupling for energy conversion, application of the idea to a molecular motor
may look straightforward.
In doing so, however, one might claim that one should construct a model
with overdamped dynamics, rather than a model with inertia.
Although it should be important to construct an overdamped-dynamics version
of the present mechanism, still the present model
may be relevant to some molecular motors, for the following reasons.

First, it is not completely sure if the model for biomotor should be
definitely overdamped.
Proteins there exhibit fluctuations of very slow time scales,
while the dynamics to realize directional motion is considerably rapid.
The time scale for the stepping motion is suggested to be faster than
microsecond order \cite{Nishiyama_Higuchi} and could be of
nanosecond order,  while the waiting time
before stepping is milliseconds.
In equilibrium, proteins show slow thermal fluctuations of the scale
of millisecond or so \cite{Wazawa_Yanagida,Yokota_Yanagida}.
Thus the time scale of the relaxation of protein might be slower than,
or of the same order as the stepping motion, in contrast with
the standard estimate.  It is still open
if the stepping of the protein motor should be definitely treated as
a overdamped motion.

It should also be noted that in the usual estimate for the damping, one uses the 
viscosity of macroscopic water, whereas
if the water around a protein can be  treated just 
as macroscopic water or not is an open question \cite{Suzuki}.  

Second, as already mentioned, each variable in our model
does not necessarily correspond to the atomic configuration  of a protein.
Rather it can correspond to some mode at a mesoscopic  scale, and
the inertia term in our model  does not directly correspond to that of an atomic motion.  
When the head swings along the chain,
microscopic  configurations do not necessarily return to the original,
but may change along the relaxation.  As long as some collective mode swings,
the present mechanism works.

A protein includes a large number of atoms.  
To understand energy transduction with such large molecule,
it is important to use a reduced model with a smaller number
of degrees of freedom representing a collective mode 
consisting of a large number of atoms.

As a theory for loose coupling, thermal ratchet models
of various forms is proposed, and they have captured some features of energy conversion.
However, some problems remain.
First, to attain a reasonable efficiency, one needs to assume a 
rather tuned timing for the change of switching of potential 
or the time scale of colored noise.  When tuned, the energy conversion
often becomes one-to-one.  Second, such external switching or a specific form of noise 
is given externally, and is not given within the theory in a
self-contained form. On the other hand, in a Feynmann-type ratchet, use of different 
temperatures is assumed, but the temperature localized at one part of a molecule
is not necessarily well defined.  Some exogenous process
is required.  Third, the ratchet
model adopts statistical description, and a single event of dynamics
is not pursued, while in the recent 
single molecule experiments, each event of
the conversion from chemical energy of ATP to mechanical work is pursued. 

Whether the concept of 'autonomous energy transducer' is valid
for biomotor or not has to be judged in experiments.
We discuss here possible predictions that can be tested experimentally
in molecular motors.
\begin{itemize}

\item
Mean step size per one ATP hydrolysis is modified by
the change of stiffness for the rail filament.
(See Fig.\ref{fig:stiffness} for the stiffness dependence of the mean step size).
Although it might be subtle which type of stiffness should be measured
to compare with our prediction, it would be important to measure
stiffness dependence in any form, as a first step.

\item
Mean step size is modified also by the change of the amount
of energy from ATP hydrolysis. If the amount of input energy is modified, the step 
size is predicted to change, roughly in proportion to it.
Although artificial modification of the available energy by ATP
is not easy experimentally, such modification of $E_0$,
might be already realized by
the long term storage of energy in molecules.
If the molecule can store the energy over several times of ATP hydrolysis,
the amount of available energy will simply increase.
Then, the step size realized in a single sequence would 
be observed to be significantly
large, compared with the usual step size of molecular motors.
In such cases, one-to-one correspondence would be lost between
ATP hydrolysis and one sequence of stepping.

\item
Molecular motors can step occasionally even without activation by ATP hydrolysis.
A rail element which has experienced stepping of motors  
accompanied by ATP hydrolysis
could possess long term memory and store a part 
of energy transduced from the motors during these events.
Then, a motor attached to this rail protein would later receive
this stored energy and use it for the stepping motion.
Similarly, stepping events without ATP may occur 
due to the energy propagated from the neighboring 
active motors with sufficient energy from ATP hydrolysis.

\item
More definite proof for the relevance of autonomous dynamics 
in the stepping motion
could be provided by direct observation of proteins dynamics.
Detection of oscillation in configuration over a few periods preceding to the stepping motion
will confirm  the validity of the autonomous dynamics to biomotors.
Measurement of conformational fluctuations with using FRETs
or with using optical traps might be effective to explore this point, 
although for complete test of the theory, rather fine time precision,
say nano or micro seconds, might be required.

\end{itemize}

To close the paper, we note that design of a nano-scale machine will be interesting
that works under the
present mechanism.  Since our scheme is general and the model is
simple, there can be several ways to realize the required situation.
With the recent advances in nano-technology, the design of autonomous 
energy transducer at a molecular scale may be realized in near future.

{\bf Acknowledgment}

We are grateful for stimulating discussion with F.Oosawa, T.Yomo, 
M.Peyrard and K.Kitamura. 
We also thank M. Peyrard for critical reading of this manuscript.
The present work is supported by
Grants-in-Aids for Scientific Research from
Japan Society for the Promotion of Science and from 
the Ministry of Education, Science and Culture of Japan.

\eject

\begin{figure}
\caption{
Profile of the model.
The form of potential $V$ depends on the value of $\theta$.
The solid curve represents the form for $\theta=0$
(the equilibrium state for $\theta$),
and the dotted curve for $\theta=\pi$.
The latter case appears by excitation,
due to the increase to $E_0$, the kinetic energy
of the pendulum.
In the simulation of this paper, the chain consists of $40$ lattice sites
with periodic boundary condition.
}
\label{fig:model}
\end{figure}

\begin{figure}
\caption{
A typical relaxation process following the excitation of the pendulum.
$K_c=0.5$, $M_{\theta}=0.01$.
In the simulation, the system was prepared in thermal equilibrium
with $T=0.01$ and $\gamma=0.01$,
and the pendulum was excited at $time=0$ with $E_0=0.4$.
At the thermal equilibrium with the temperature here, the head remains
at one lattice site over a very long time.
Upper: Time series of the positions of the head $x_h$ (solid line)
and a few neighboring lattice sites $x_i$ ($-1\le i\le 3$, dotted lines).
Lower: Time series of the energy of the internal pendulum
$E_{\theta}=T_{\theta}+V_{\theta}$ ($0\le E_{\theta} \le E_0$),
kinetic energy of each lattice site $T_i$ and the head $T_h$
($0\le T_i \le 0.16$) are plotted, where the vertical scale
of $T_{\theta}$ is larger than those for the others.
}
\label{fig:jikeiretsu}
\end{figure}

\begin{figure}
\caption{
(a) Mean displacement $\langle \Delta x_h \rangle$ of the head
for four values of injected energy $E_0$,
$0$ (equilibrium), $0.1$, $0.3$ and $0.5$. $K_c=0.5$. $\gamma=0.01$.
(b) Diffusion coefficient of the head along  the chain in thermal
equilibrium with $\gamma=0.01$. Here $D\equiv \langle (\Delta x_h-\langle \Delta x_h\rangle )^2\rangle /\Delta t$. $\Delta t =1000$.
}
\label{fig:temperature}
\end{figure}

\begin{figure}
\caption{
Average displacement $\langle \Delta x_h\rangle$ as a function of
$E_0$ for three values of temperature,
computed as the ensemble average over 1000 samples. $K_c=0.5$.
For larger values of $E_0$,
$\langle \Delta x_h\rangle$ increases as a function of $E_0$
similarly for all $T$.
}
\label{fig:injected_energy}
\end{figure}

\begin{figure}
\caption{
Dependence of average displacement $\langle \Delta x_h\rangle$ on $K_c$
for two temperature $T=0.01, 0.02$ with $E_0=0.4$.
Directional motion is most prominent in a particular range of
stiffness of the chain ($K_c \approx 0.5$), where
the fluctuations of the lattice are slightly larger than those of the head.
The directional motion is suppressed both for larger values of $K_c$ and
smaller values of $K_c$.  The former decrease is due to the decrease pf
the magnitude of lattice fluctuations, while
the latter decrease is
due to the interaction of the head not only with the adjacent lattice site,
but also with neighboring sites, resulting in an effectively
stronger potential experienced by the head.
}
\label{fig:stiffness}
\end{figure}

\begin{figure}
\caption{
Sensitivity of the dynamics to small disturbances.
The small difference at the event of excitation induces
very large difference of the motion of the head $x_h$.
In the simulation, the system was prepared in thermal equilibrium
before excitation with $T=0.01$ and $\gamma=0.01$,
and the pendulum was excited at $time=0$ with $E_0=0.35$.
At the same time, small disturbance of the order $10^{-10}$ is added
to the value of $x_h$.
Initial conditions and applied random sequence are common to the two
time sequence.
}
\label{fig:difference}
\end{figure}

\begin{figure}
\caption{
(a) Frequency distribution of the displacement (number of steps)
of the head position $x_h$ per excitation.
$K_c=0.5$, $T=0.006$ and $E_0=0.5$.
(b) Relation between the mean and the variance for the step distribution
in (a).
$K_c=0.5$, $\gamma=0.01$ for three kinds of temperatures.
The graph displays data for various values of $E_0$.
}
\label{fig:distribution}
\end{figure}

\begin{figure}
\caption{
Time sequence of $W_{in}$ and $W_{out}$ for
(a) $K_c=0.5$, $T=0.01$, $E_0=0.5$ and $m_{\theta}=0.02$,
and (b) $K_c=1.0$, $T=0.01$, $E_0=0.5$ and $m_{\theta}=0.02$.
Lower figures are corresponding time series of the position $x_h$
for the head. For detailed explanation, see the text.
}
\label{fig:flow}
\end{figure}

\begin{figure}
\caption{
Time sequence for (a) the mean flow rate $d\langle W_{out}\rangle /dt$ and
(b) the mean sustained energy $\langle W_{in}-W_{out}\rangle$ in the active part,
for $K_c=0.5$ (solid line) and $K_c=1.3$ (dotted line).
$E_0=0.35$ and $m_h=0.02$. In this simulation, lattice sites
satisfying $i\le i_0-3$ and $i\ge i_0+3$ are contacted with the heat bath with
$\gamma=0.05$ and $T=0.01$,  in order to make the relation in energy flow
clear around the active part.
 }
\label{fig:mean_flow}
\end{figure}

\begin{figure}
\caption{
The rate $\langle W_{in}-W_{out} \rangle/\langle W_{in} \rangle$,
at the mean waiting time $t_w$.
The contact with the heat bath is given in the same manner
as in Fig.{\protect \ref{fig:mean_flow}}, where $\gamma=0.05$ and $T=0.01$.
The rate is almost constant over a wide range of $K_c$
for small injected energy $E_0=0.25$, which is smaller than the threshold energy
for the directional motion even for $K_c \approx 0.5$
(see Fig.{\protect\ref{fig:injected_energy}}).
For larger values of $E_0$, the rate depends strongly on $K_c$.
Approximately half of the energy
flowing into the active part is sustained even though the chain is very stiff.
This is because $t_w$ is as short time as $100$. Indeed, if we choose larger $t_w$,
this rate becomes smaller.
}
\label{fig:maintained_energy}
\end{figure}

\begin{figure}
\caption{
Frequency distribution for the extracted energy $W_{in}(t_n)$
from the pendulum up to $n$th step.
$W_{in}(t_n)$ is measured at the time when  the head crosses $n$th 
energy barrier to shift from $n-1$th to $n$th site.
$T=0.01$, $\gamma=0.01$, $E_0=0.4$, $m_h=0.02$  and $K_c=0.6$.
}
\label{fig:degree_by_degree}
\end{figure}

\begin{figure}
\caption{
Dependence of the distributions of the rail distributions upon imposed load.
Frequency distributions of $\Delta x_0$ is plotted, by using
the configuration of the system described in the text.
$T=0.01$, $\gamma=0.01$, $E_0=0.6$, $K_c=0.5$ and $m_h=0.02$.
The solid and dotted lines give distributions for a short ($N_c=10$)
and a longer rail ($N_c=200$), respectively.
}
\label{fig:load-distribution}
\end{figure}

\end{document}